# Magnon-mediated spin currents in Tm$_3$Fe$_5$O$_{12}$/Pt with perpendicular magnetic anisotropy

G. L. S. Vilela[1,2], J. E. Abrao[3], E. Santos[3], Y. Yao[4,5], J. B. S. Mendes [6], R. L. Rodríguez-Suárez[7], S. M. Rezende[3], W. Han[4,5], A. Azevedo[3], and J. S. Moodera[1,8]

[1] Plasma Science and Fusion Center, and Francis Bitter Magnet Laboratory, Massachusetts Institute of Technology, Cambridge, Massachusetts 02139, USA
[2] Física de Materiais, Escola Politécnica de Pernambuco, Universidade de Pernambuco, Recife, Pernambuco 50720-001, Brasil
[3] Departamento de Física, Universidade Federal de Pernambuco, Recife, Pernambuco 50670-901, Brasil
[4] International Center for Quantum Materials, School of Physics, Peking University, Beijing 100871, China.
[5] Collaborative Innovation Center of Quantum Matter, Beijing 100871, China.
[6] Departamento de Física, Universidade Federal de Viçosa, Viçosa, Minas Gerais 36570-900, Brasil
[7] Facultad de Física, Pontificia Universidad Católica de Chile, Casilla 306, Santiago, Chile
[8] Department of Physics, Massachusetts Institute of Technology, Cambridge, Massachusetts 02139, USA

Electronic mail: gilvania.vilela@upe.br

**Abstract**

The control of pure spin currents carried by magnons in magnetic insulator (MI) garnet films with a robust perpendicular magnetic anisotropy (PMA) is of great interest to spintronic technology as they can be used to carry, transport and process information. Garnet films with PMA present labyrinth domain magnetic structures that enrich the magnetization dynamics, and could be employed in more efficient wave-based logic and memory computing devices. In MI/NM bilayers, where NM being a normal metal providing a strong spin-orbit coupling, the PMA benefits the spin-orbit torque (SOT) driven magnetization's switching by lowering the needed current and rendering the process faster, crucial for developing magnetic random-access memories (SOT-MRAM). In this work, we investigated the magnetic anisotropies in thulium iron garnet (TIG) films with PMA via ferromagnetic resonance measurements, followed by the excitation and detection of magnon-mediated pure spin currents in TIG/Pt driven by microwaves and heat currents. TIG films presented a Gilbert damping constant $\alpha \approx 0.01$, with resonance fields above 3.5 kOe and half linewidths broader than 60 Oe, at 300 K and 9.5 GHz. The spin-to-charge current conversion through TIG/Pt was observed as a micro-voltage generated at the edges of the Pt film. The obtained spin Seebeck coefficient was 0.54 $\mu V/K$, confirming also the high interfacial spin transparency.

Spin-dependent phenomena in systems composed by layers of magnetic insulators (MI) and non-magnetic heavy metals (NM) with strong spin-orbit coupling have been extensively explored in the insulator-based spintronics [1-6]. Among the MI materials, YIG (Y$_3$Fe$_5$O$_{12}$) is widely employed in devices for generation and transmission of pure spin currents. The main reason is its very low magnetic damping with Gilbert parameter on the order of 10$^{-5}$, and its large spin decay length which permits spin waves to travel distances of orders of centimeters inside it before they vanish [7-9]. When combined with heavy metals such as Pt, Pd, Ta, or W, many intriguing spin-current related phenomena emerge, such as the spin pumping effect (SPE) [10-14], spin Seebeck effect (SSE) [7, 15-18], spin Hall effect (SHE) [19-21], and spin-orbit torque (SOT) [22-25]. The origin of these effects relies mainly on the spin diffusion length, and the quantum-mechanical exchange and spin-orbit interactions at the interface and inside the heavy metal [26]. All of these effects turn out the MI/NM bilayer into a fascinating playground for exploring spin-orbit driven phenomena at interfaces [27-30].

Well investigated for many years, intrinsic YIG(111) films on GGG(111), (GGG = Gd$_3$Ga$_5$O$_{12}$) exhibit in-plane anisotropy. To obtain YIG single-crystal films with perpendicular magnetic anisotropy (PMA) it is necessary to grow them on top of a different substrate or partially substitute the yttrium ions by rare-earth ions, to cause strain-induced anisotropy [31-33]. Even so, it is well-known that magnetic films with PMA play an important role in spintronic technology. The PMA enhances the spin-switching efficiency, which reduces the current density for observing the spin-orbit torque (SOT) effect, and it is useful for developing SOT based magnetoresistive random access memory (SOT-MRAM) [34-36]. Besides that, PMA increases the information density in hard disk drives and magnetoresistive random access memories [37-39], and it is crucial for breaking the time-reversal symmetry in topological insulators (TIs) aiming towards quantized anomalous Hall state in MI/TI [40-42].

Recently, thin films of another rare-earth iron garnet, TIG (Tm$_3$Fe$_5$O$_{12}$), have caught the attention of researchers due to its large negative magnetostriction constant, which favors an out-of-plane easy axis [4, 43, 44]. TIG is a ferrimagnetic insulator with a critical temperature of 549 K, a crystal structure similar to YIG, and a Gilbert damping parameter on the order of $\alpha \sim 10^{-2}$ [4, 45]. Investigations of spin transport effects have been reported in TIG/Pt [45, 46] and TIG/TI [42, 47], where the TIG was fabricated by pulsed laser deposition (PLD) technique. The results showed a strong spin mixing conductance at the interface of these materials that made it possible to observe spin Hall magnetoresistance, spin Seebeck, and spin-orbit torque effects.

In this paper, we first present a study of the magnetocrystalline and uniaxial anisotropies, as well as the magnetic damping of sputtered epitaxial TIG thin films using the ferromagnetic resonance (FMR) technique. For obtaining the cubic and uniaxial anisotropy fields, we analyzed the dependence of the FMR spectra on the film thickness and the orientation of the dc applied magnetic field at room temperature and 9.5 GHz. Then, we swept the microwave frequency for getting their magnetic damping at different temperatures. Subsequently, we focused this investigation on the excitation of magnon-mediated pure spin currents in TIG/Pt via the spin pumping and spin Seebeck mechanisms for different orientations of the dc applied magnetic field at room temperature. Pure spin currents transport spin angular momentum without carrying charge currents. They are free of Joule heating and could lead to spin-wave based devices that are energetically more efficient. Employing the inverse spin Hall effect (ISHE) [12], we observed the spin-to-charge conversion of these currents inside the Pt film which was detected as a developed micro-voltage.

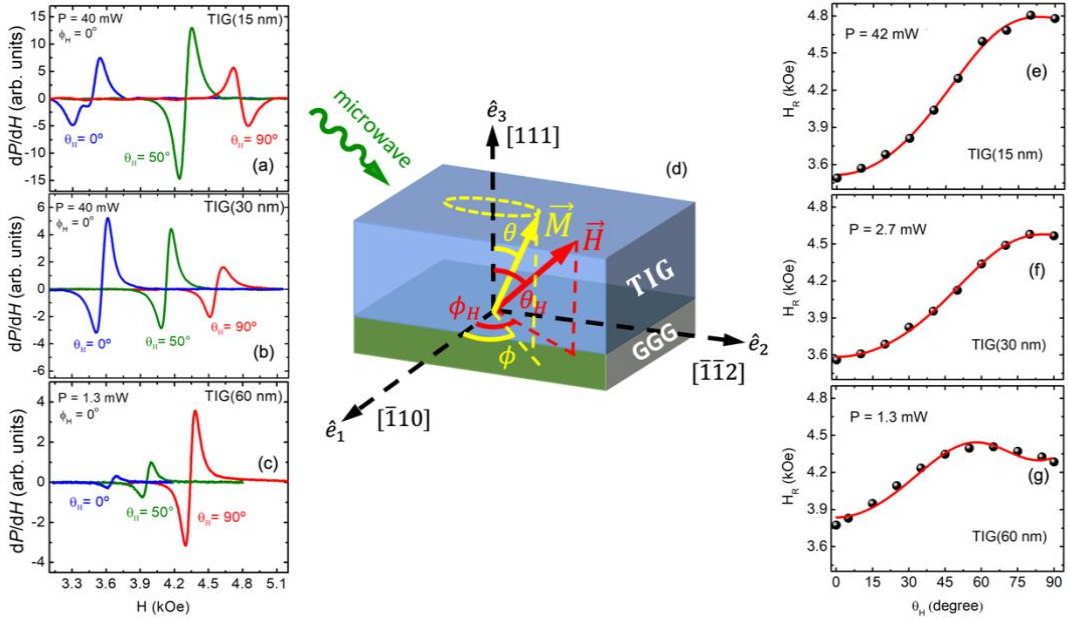

**Fig. 1.** FMR absorption derivative spectra vs. field scan H for (a) TIG(15 nm), (b) TIG(30 nm), and (c) TIG(60 nm), at room T and 9.5 GHz. The half linewidths ($\Delta H$) for TIG(15 nm) with $H$ applied along $\theta_H = 0°, 50°$, and $90°$ are 112 Oe, 74 Oe, and 72 Oe, respectively. For TIG(30 nm), $\Delta H$ is 82 Oe, 72 Oe, and 65 Oe for $\theta_H = 0°, 50°$, and $90°$, respectively. For TIG(60 nm), $\Delta H$ is 72 Oe, 75 Oe, and 61 Oe for $\theta_H = 0°, 45°$, and $90°$, respectively. These values were extracted from the fits using the Lorentz function. (d) Illustration of the FMR experiment where the magnetization ($M$) under an applied magnetic field (H) is driven by a microwave. (e), (f) and (g) show the dependence of the resonance field $H_R$ with $\theta_H$ for different thickness of TIG. The red solid lines are theoretical fits obtained for the FMR condition. Magnetization curves are given in reference [44].

TIG films with thickness ranging from 15 to 60 nm were deposited by rf sputtering from a commercial target with the same nominal composition of $Tm_3Fe_5O_{12}$, and a purity of 99.9 %. The deposition process was performed at room temperature, in pure argon working pressure of 2.8 mTorr, at a deposition rate of 1.4 nm/min. To improve the crystallinity and the magnetic ordering, the films were post-growth annealed for 8 hrs at 800 °C in a quartz tube in flowing oxygen. After the thermal treatment, the films yielded a magnetization saturation of 100 emu/cm³, and an RMS roughness below 0.1 nm confirmed using a superconducting quantum interference device (SQUID) and high-resolution X-ray diffraction measurements, as detailed in our recent article [44]. Moreover, the out-of-plane hysteresis loops showed curved shapes which might be related with labyrinth domain structures very common in garnet films with PMA [48]. The next step of sample preparation consisted of an ex-situ deposition of a 4 nm-thick Pt film over the post-annealed TIG films using the dc sputtering technique. Platinum films were grown under an Ar gas pressure of 3.0 mTorr, at room temperature, and a deposition rate of 10 nm/min. The Pt films were not patterned.

Ferromagnetic resonance (FMR) is a well-established technique for study of basic magnetic properties such as saturation magnetization, anisotropy energies and magnetic relaxation mechanisms. Furthermore, FMR has been central to the investigation of microwave-driven spin-pumping phenomena in FM/NM bilayers [11, 12, 49]. First, we used a homemade FMR spectrometer running at a fixed frequency of 9.5 GHz, at room temperature, where the samples were placed in the middle of the back wall of a rectangular microwave cavity operating in the $TE_{102}$ mode with a Q factor of 2500. Field scan spectra of the derivative of the absorption power ($dP/dH$) were acquired by modulating the dc applied field $\vec{H}_0$ with a small sinusoidal field $\vec{h}$ at 100 kHz and using lock-in amplifier detection. The resonance field $H_R$ was obtained as a function of the polar and azimuthal angles ($\theta_H, \phi_H$) of the applied magnetic field $\vec{H}$, as illustrated in Fig. 1(d), where $\vec{H} = \vec{H}_0 + \vec{h}$ and $h \ll H_0$.

The FMR spectra for TIG(t) films are shown in Figs. 1 (a, b and c) for thicknesses t = 15, 30 and 60 nm respectively. The spectra were measured for $H$ applied along three different polar angles: $\theta_H = 0°$ (blue), $\theta_H \cong 45°$ (green) and $\theta_H = 90°$ (red). The complete dependence of $H_R$, for each sample, as a function of the polar angle (0° ≤ $\theta_H$ ≤ 90°) are shown in Figs. 1(e, f and g). For all samples, $H_R$ was minimum for $\theta_H$ = 0°, confirming that the perpendicular anisotropy field was strong enough to overcome the demagnetization field. While the films with t = 15 nm and 30 nm exhibited the maximum value of $H_R$ for $\theta_H$ = 90° (in-plane), the sample with t = 60 nm showed a maximum $H_R$ at $\theta_H$ ∼60°. To explain the behavior of $H_R$ as a function of the out-of-plane angle $\theta_H$, it is necessary to normalize the FMR data to compare with the theory described as follows.

The most relevant contributions to the free magnetic energy density $\epsilon$ for GGG(111)/ TIG(111) films, are:

$$\epsilon = \epsilon_Z + \epsilon_{CA} + \epsilon_D + \epsilon_U, \quad (1)$$

where $\epsilon_Z$ is the Zeeman energy density, $\epsilon_{CA}$ is the cubic anisotropy energy density for (111) oriented thin films, $\epsilon_D$ is the demagnetization energy density, and $\epsilon_U$ is the uniaxial energy density. Taking into consideration the reference frame shown in Fig. 1(d), each energy density terms can be written as [50]:

$$\epsilon_Z = -M_S H(\sin\theta \sin\theta_H \cos(\phi - \phi_H) + \cos\theta \cos\theta_H), \quad (2)$$

$$\epsilon_{CA} = K_1/12 \left(3 - 6\cos^2\theta + 7\cos^4\theta + 4\sqrt{2}\cos\theta \sin 3\phi \sin^3\theta\right), \quad (3)$$

$$\epsilon_D + \epsilon_U = 2\pi(\vec{M}\cdot\hat{e}_3)^2 - K_2^\perp(\vec{M}\cdot\hat{e}_3/M_S)^2 - K_4^\perp(\vec{M}\cdot\hat{e}_3/M_S)^4, \quad (4)$$

where $\theta$ and $\phi$ are the polar and azimuthal angles of the magnetization vector $\vec{M}$, $M_S$ is the saturation magnetization, $K_1$ is the first order cubic anisotropy constant, and $K_2^\perp$ and $K_4^\perp$ are the first and second order uniaxial anisotropy constants. The uniaxial anisotropy terms come from two sources: growth induced and stress induced anisotropy. The relation between the resonance field and the excitation frequency $\omega$ can be obtained from [51, 52]:

| TIG film's thickness t | 15 nm | 30 nm | 60 nm |
|---|---|---|---|
| $4\pi M_{eff}$ (G) | -979 | -799 | -383 |
| $H_{1C} = 2K_1/M_S$ (Oe) | 31 | 26 | -111 |
| $H_{U2} = 4\pi M_{eff} - 4\pi M_S$ (Oe) | -2,739 | -2,559 | -2,143 |
| $H_{U4} = 2K_4^\perp/M_S$ (Oe) | 311 | 168 | 432 |

**Table 1.** Physical parameters extracted from the theoretical fits of the FMR response of the TIG thin films with thickness $t$, performed at room $T$ and 9.5 GHz. $4\pi M_{eff}$ is the effective magnetization, $H_{1C}$ is the cubic anisotropy field, $H_{U2}$ and $H_{U4}$ are the first and second order uniaxial anisotropy fields, respectively. $H_{U2}$ is the out-of-plane uniaxial anisotropy field, also named as $H_\perp$.

$$(\omega/\gamma)^2 = \frac{1}{M^2 \sin^2\theta}\left[\epsilon_{\theta\theta}\epsilon_{\phi\phi} - (\epsilon_{\theta\phi})^2\right], \quad (5)$$

where $\gamma$ is the gyromagnetic ratio. The subscripts indicate partial derivatives with respect to the coordinates, $\epsilon_{\theta\theta} = \partial^2\epsilon/\partial\theta^2|_{\theta_0,\phi_0}$, $\epsilon_{\phi\phi} = \partial^2\epsilon/\partial\phi^2|_{\theta_0,\phi_0}$ and $\epsilon_{\theta\phi} = \partial^2\epsilon/\partial\theta\partial\phi|_{\theta_0,\phi_0}$, where $\theta_0, \phi_0$ are the equilibrium angles of the magnetization determined by the energy density minimum conditions, $\partial\epsilon/\partial\theta|_{\theta_0,\phi_0} = 0$ and $\partial\epsilon/\partial\phi|_{\theta_0,\phi_0} = 0$. The best fits to the data obtained with the Eq. (5) are shown in Figs. 1 (e, f and g) by the solid red lines. The main physical parameters extracted from the fits, including the effective magnetization $4\pi M_{eff}$, are summarized in Table 1. Here $4\pi M_{eff} = 4\pi M - 2K_2^\perp/M_S$, where the second term is the out-of-plane uniaxial anisotropy field $H_{U2} = 2K_2^\perp/M$, also named as $H_\perp$. It is important to notice that the large negative values of $H_{U2}$ were sufficiently strong to saturate the magnetization along the direction perpendicular to the TIG film's plane, thus overcoming the shape anisotropy. We used the saturation magnetization as the nominal value of $M_S = 140.0\ G$. As the thickness of the TIG film increased, the magnitude of the perpendicular magnetic anisotropy field, $H_{U2}$, decreased due to the relaxation of the induced growth stresses as expected.

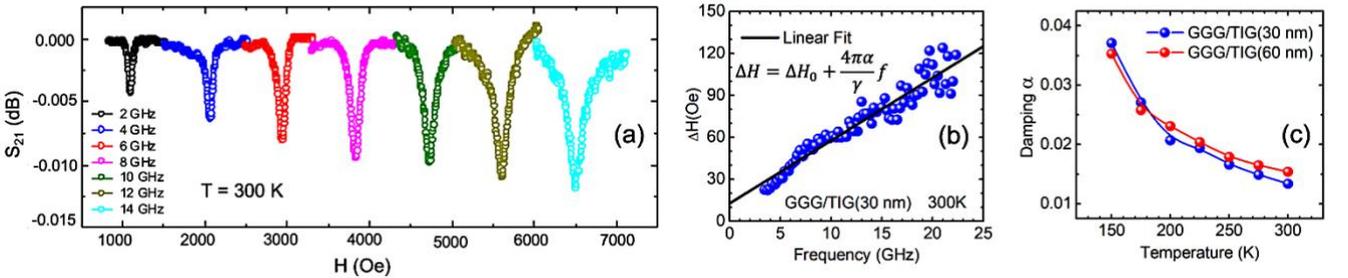

**Fig. 2.** (a) Ferromagnetic resonance spectra vs. in-plane applied field $H$ for a 30 nm-thick TIG film at frequencies ranging from 2 GHz to 14 GHz and temperature of 300 K, after normalization by background subtraction. (b) Half linewidth $\Delta H$ versus frequency for TIG(30 nm) at 300 K. The Gilbert damping parameter $\alpha$ was extracted from the linear fitting of the data. (c) Damping $\alpha$ versus temperature $T$ for TIG films with 30 nm and 60 nm of thickness.

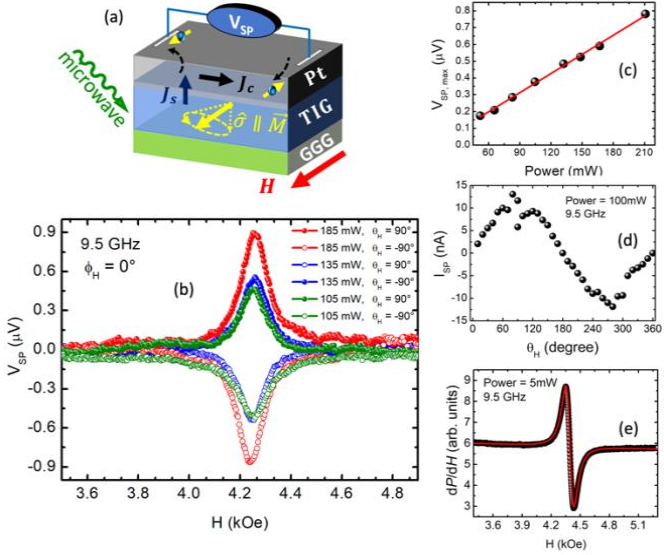

for different microwave frequencies ($f$) and temperatures ($T$). Figure 2(a) shows the FMR spectra ($S_{21}$ versus $H$) for TIG(30 nm) corresponding to frequencies ranging from 2 GHz to 14 GHz at 300K, with a microwave power of 0 dBm, after normalization by background subtraction. Fitting each FMR spectra using the Lorentz function, we were able to extract the half linewidth $\Delta H$ for each frequency, as shown in Fig. 2(b). Then, $\alpha$ was estimated based on the linear approximation $\Delta H = \Delta H_0 + (4\pi\alpha/\gamma)f$, where $\Delta H_0$ reflects the contribution of magnetic inhomogeneities, the linear frequency part is caused by the intrinsic Gilbert damping mechanism, and $\gamma$ is the gyromagnetic ratio [40]. The same analysis was performed for lower temperature data, and it was extended to TIG(60 nm). Due to the weak magnetization of the thinnest TIG (15 nm) the coplanar waveguide setup was not able to detect its FMR signals. Figure 2(c) shows the Gilbert damping dependence with $T$. At 300 K, $\alpha = 0.015$ for TIG(60 nm) which is in agreement with the values reported in the literature [4, 45], and it increases by 130 % as $T$ goes down to 150 K [54].

Next, this work focused on the generation of pure spin currents carried by spin waves in TIG at room $T$, followed by their propagation through the interface between TIG and Pt, and their spin-to-charge conversion inside the Pt film. Initially, we explored the FMR-driven spin-pumping effect in TIG(60 nm)/Pt(4 nm), where the coherent magnetization precession of the TIG injected a pure spin current $J_s$ into the Pt layer, which converted as a transverse charge current $J_c$ by means of the inverse spin Hall effect, expressed as $\vec{J}_c = \theta_{SH}(\hat{\sigma} \times \vec{J}_s)$, where $\theta_{SH}$ is the spin Hall angle and $\hat{\sigma}$ is the spin polarization [55]. As the FMR was excited using the homemade spectrometer at 9.5 GHz, a spin pumping voltage ($V_{SP}$) was detected between the two silver painted electrodes

**Fig. 3.** Spin pumping voltage ($V_{SP}$) excited by a FMR microwave of 9.5 GHz, at room $T$, in TIG(60 nm)/Pt(4 nm). (a) Illustration of the spin pumping setup. (b) In-plane field scan of $V_{SP}$ for different microwave powers. (c) Linear dependence of the maximum $V_{SP}$ with the microwave power. (d) $\theta_H$ scan of the charge current ($I_{SP}$) generated by means of the inverse spin Hall effect in the Pt film. (e) In-plane field scan of the FMR absorption derivative spectrum for 5 mW.

To obtain the Gilbert damping parameter ($\alpha$) of the TIG thin films, we used the coplanar waveguide technique in the variable temperature insert of a physical property measurement system (PPMS). A vector network analyzer measured the amplitude of the forward complex transmission coefficients ($S_{21}$) as a function of the in-plane magnetic field placed on the edges of the Pt film, as illustrated in Fig. 3(a). It is important to note that when the magnetization vector was perpendicular to the sample's plane no $V_{SP}$ was detected. The sample TIG(60 nm)/Pt(4 nm) had dimensions of 3 x 4 mm$^2$, and a resistance between the silver electrodes of **48 Ω** at zero field. The $V_{SP}$ showed a peak value of $\mathbf{0.85\ \mu V}$ in the resonance magnetic field for an incident power of 185 mW, and an in-plane dc magnetic field ($\theta_H = 90°$) as shown in Fig. 3(b). The signal reversed when the field direction went through a $\mathbf{180°}$ rotation. The dependence of $V_{SP}$ with the microwave incident power was linear, as shown in Fig. 3(c), whereas the spin pumping charge current ($I_{SP} = V_{SP}/R$) had the dependence of $V_{SP} \propto \sin\theta_H$, showed in Fig. 3(d), for a fixed microwave power of 100 mW. The ratio between the microwave-driven voltage and the microwave power was 4μV/W.

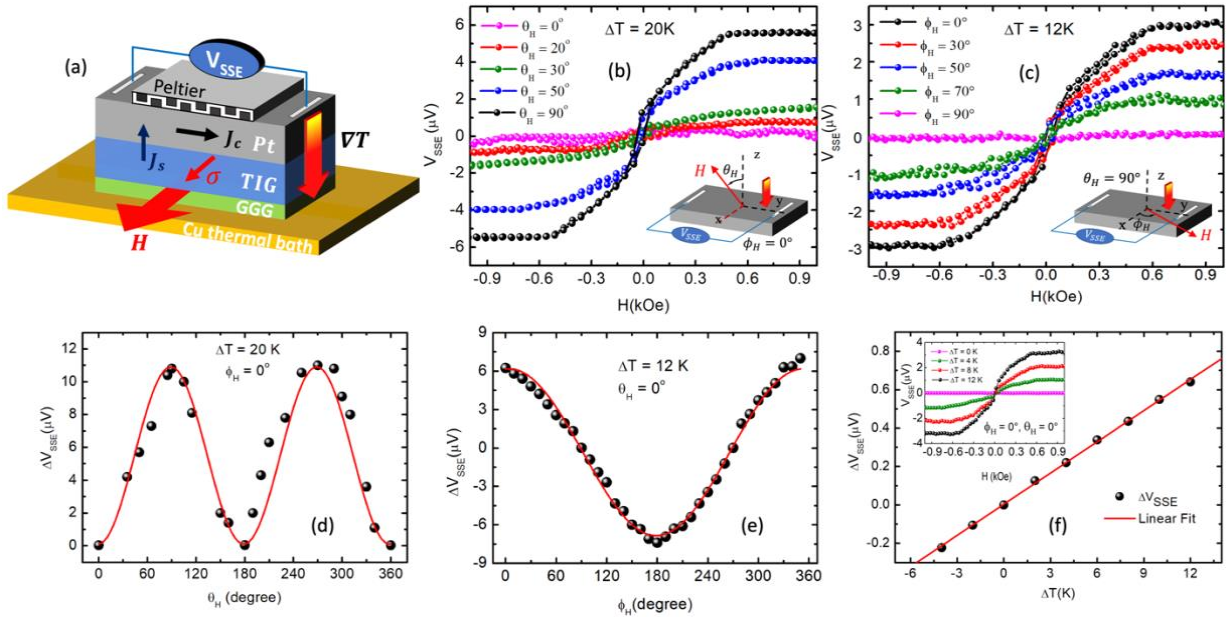

**Fig. 4.** Spin Seebeck voltage ($V_{SSE}$) excited by a thermal gradient in the longitudinal configuration ($\nabla T \parallel \vec{J}_S$) at room $T$, as shown in (a). (b) Field scan of $V_{SSE}$ for $\Delta T = 20K$ and different field polar angles $\theta_H$. (c) Field scan of $V_{SSE}$ for $\Delta T = 12$, $\theta_H = 90°$ and different azimuthal angles $\phi_H$. Spin voltage amplitude $\Delta V_{SSE}$ versus (d) $\theta_H$, (e) $\phi_H$, and (f) $\Delta T$. The solid red lines are theoretical fits of the sine (d), cosine (e) and linear (f) dependence of $\Delta V_{SSE}$ with $\theta_H$, $\phi_H$ and $\Delta T$, respectively.

We also excited pure spin currents via the spin Seebeck effect (SSE) in TIG(60 nm)/Pt(4 nm) at room $T$. SSE emerges from the interplay between the spin and heat currents, and it has the potential to harvest and reduce power consumption in spintronic devices [16, 18]. When a magnetic material is subjected to a temperature gradient, a spin current is thermally driven into the adjacent non-magnetic (NM) layer by means of the spin-exchange interaction. The spin accumulation in the NM layer can be detected by measuring a transversal charge current due to the ISHE. To observe the SSE in our samples, the uncovered GGG surface was placed over a copper plate, acting as a thermal bath at room $T$, while the sample's top was in thermal contact with a $\mathbf{2 \times 2\ mm^2}$ commercial Peltier module through a thermal paste, as illustrated in Fig. 4(a). The Peltier module was responsible for creating a controllable temperature gradient across the sample. On the other hand, the temperature difference ($\Delta T$) between the bottom and top of the sample was measured by a differential thermocouple. The ISHE voltage due to the SSE ($V_{SSE}$) was detected between the two silver painted electrodes placed on the edges of the Pt film.

The behaviour of $V_{SSE}$ by sweeping the dc applied magnetic field ($H$), while $\Delta T$, $\theta_H$ and $\phi_H$ were kept fixed was investigated. Fixing $\phi_H = 0°$ and varying the magnetic field from out-of-plane ($\theta_H = 0°$) to in-plane along x-direction ($\theta_H = 90°$), $V_{SSE}$ went from zero to its maximum value of 5.5 $\mu V$ for $\Delta T = 20\ K$, as shown in Fig. 4(b). Around zero field, no matter the value of $\theta_H$, the TIG's film magnetization tended to rely along its out-of-plane easy axis which zeroes $V_{SSE}$. For in-plane fields ($\theta_H = 90°$) with $\Delta T = 12\ K$, $V_{SSE}$ was maximum when $\phi_H = 0°$, and it was zero for $\phi_H = 90°$. The reason $V_{SSE}$ went to zero for $\phi_H = 90°$, may be attributed to the generated charge flow along the x-direction while the silver electrodes were placed along y-direction, thus not enabling the current detection (see Fig. 4(c)). The analysis of the spin Seebeck amplitude $\Delta V_{SSE}$ versus $\theta_H$, $\phi_H$ and $\Delta T$ showed a sine, cosine and linear dependence, respectively as can be seen in Fig. 4(d)-(e), where the red solid lines are theoretical fits. The Spin Seebeck coefficient (SSC) extracted from the linear fit of $\Delta V_{SSE}$ vs. $\Delta T$ was 0.54 $\mu V/K$.

In conclusion, we used the FMR technique to probe the magnetic anisotropies and the Gilbert damping parameter of the sputtered TIG thin films with perpendicular magnetic anisotropy. The results showed higher resonance fields (> 3.5 kOe) and broader linewidths (> 60 Oe) when comparing with YIG films at room $T$. Thinner TIG films (t = 15 nm and 30 nm) presented a well-defined PMA; on the other hand, the easy axis of thicker TIG film (60 nm) showed a deviation of 30 degrees from normal to the film plane. By numerically adjusting the FMR field dependence with the polar angle, we extracted the effective magnetization, the cubic ($H_{1C}$) and the out-of-plane uniaxial anisotropy ($H_{U2} = H_\perp$) fields for the three TIG films. The thinnest film presented the highest intensity for $H_\perp$ as expected, even so $H_\perp$ was strong enough to overcome the shape anisotropy and gave place to a perpendicular magnetic anisotropy in all the three thickness of TIG films. The Gilbert damping parameter ($\alpha$) for TIG(30 nm) and TIG(60 nm) films were estimated to be ≈ 10$^{-2}$, by analyzing a set of FMR spectra using the coplanar waveguide technique at various microwave frequencies and temperatures. As $T$ went down to 150 K the damping increased monotonically 130 %.

Furthermore, spin waves (magnons) were excited in TIG(60 nm)/Pt(4 nm) heterostructure through the spin pumping and spin Seebeck effects, at room $T$ and 9.5 GHz. The generated pure spin currents carried by the magnons were converted into charge currents once they reached the Pt film by means of the inverse spin Hall effect. The charge currents were detected as a micro-voltage measured at the edges of the Pt film, and they showed sine and cosine dependence with the polar and azimuthal angles, respectively, of the dc applied magnetic field. This voltage was linearly dependent on the microwave power for the SPE, and on the temperature gradient for the SSE. These results confirmed a good spin-mixing conductance in the interface TIG/Pt, and an efficient conversion of pure spin currents into charge currents inside the Pt film, which is crucial for the employment of TIG films with a robust PMA in the development of magnon-based spintronic devices for computing technologies.

## ACKNOWLEDGEMENTS

This research is supported in the USA by Army Research Office (ARO W911NF-19-2-0041 and W911NF-19-2-0015), NSF (DMR 1700137), ONR (N00014-16-1-2657), in Brazil by CAPES (Gilvania Vilela/POS-DOC-88881.120327/2016-01), FACEPE (APQ-0565-1.05/14 and APQ-0707-1.05/14), CNPq, UPE (PFA/PROGRAD/UPE 04/2017) and FAPEMIG - Rede de Pesquisa em Materiais 2D and Rede de Nanomagnetismo, in Chile by Fondo Nacional de Desarrollo Científico y Tecnológico (FONDECYT) No. 1170723, and in China by the National Natural Science Foundation of China (11974025).

## DATA AVAILABILITY

The data that support the findings of this study are available from the corresponding author upon reasonable request.